\pdfoutput=1
\PassOptionsToPackage{pdftex}{graphicx}
\documentclass[twocolumn,letter]{jpsj3}
\usepackage{txfonts,textcomp}

\begin{document}

\title{Superconductivity at the Pressure-Induced Ferromagnetic
Critical Region in UCoGe}

\author{Masahiro Manago$^{1,2}$\thanks{manago@crystal.kobe-u.ac.jp},
Shunsaku Kitagawa$^1$,
Kenji Ishida$^1$,
Kazuhiko Deguchi$^3$,\\
Noriaki K. Sato$^3$,
Tomoo Yamamura$^4$\thanks{Present address: Institute for Integrated Radiation
and Nuclear Science, Kyoto University, Kumatori 590-0494, Japan}
}
\inst{
$^1$Department of Physics, Graduate School of Science,
Kyoto University, Kyoto 606-8502, Japan\\
$^2$Department of Physics, Graduate School of Science,
Kobe University, Kobe 657-8501, Japan\\
$^3$Department of Physics, Graduate School of Science,
Nagoya University, Nagoya 464-8602, Japan\\
$^4$Institute for Materials Research,
Tohoku University, Sendai 980-8577, Japan
}

\abst{
The phase separation of the ferromagnetic (FM) and paramagnetic (PM) phases in
the superconducting (SC) state of UCoGe at the FM critical region was
investigated using $^{59}$Co nuclear quadrupole resonance (NQR) technique
by taking advantage of its site-selective feature.
The NQR measurements revealed that the first-order quantum phase
transition occurs between the FM and the PM states.
The nuclear spin-lattice relaxation rate $1/T_1$ exhibited a clear drop at the
SC state in the PM phase, whereas it was not detected in the FM phase, which
indicates that the superconductivity in the FM phase becomes weaker at the
FM critical region due to the presence of the PM SC state.
This result suggests that the SC condensation energy of the PM SC state is equal
or larger than that of the FM SC state in this region.
The pressure--temperature phase diagram of UCoGe was modified by taking
the results from this study into account.
}

\maketitle

Uranium (U)-based ferromagnetic (FM) superconductors have attracted much
attention since the discovery of superconductivity in UGe$_2$ under pressure
\cite{Nature.406.587} and ambient-pressure superconductivity in URhGe
\cite{Nature.413.6856} and UCoGe \cite{PhysRevLett.99.067006}.
The spin-triplet superconductivity is expected in these systems and
is attributed to FM spin fluctuations.
The pairing glue is tuned by the external magnetic field; therefore,
the superconducting (SC) phases of these compounds exhibit
anomalous field--temperature phase diagrams
including re-entrant superconductivity
\cite{PhysRevB.64.220503,Science.309.1343,JPSJ.78.113709}.
These U-based superconductors are the key materials
for understanding the FM quantum criticality.
The FM phase is tuned by the application of pressure and magnetic field.
The SC phase exists only inside the FM phase in UGe$_2$
\cite{Nature.406.587}, whereas it remains in the paramagnetic (PM) state in
UCoGe \cite{JPSJ.77.073703,PhysRevLett.103.097003,PhysRevB.94.125110}.
The pressure phase diagram of UCoGe is similar to that of
the materials exhibiting superconductivity near a quantum critical point.
Some studies reveal that a tricritical point in the field--temperature diagram
of URhGe plays an important role for the re-entrant superconductivity
\cite{PhysRevLett.114.216401,PhysRevB.93.201112}.
The spin fluctuations in URhGe differ from the simple Ising-type ones
as observed in UCoGe\cite{PhysRevLett.105.206403,PhysRevLett.108.066403} and
UGe$_2$ \cite{JPSJ.87.033704}.

Recently, we reported an enhancement of the FM fluctuations in UCoGe
caused by the suppression of the FM phase by the pressure, as revealed
by the $^{59}$Co
nuclear quadrupole resonance (NQR) measurement\cite{PhysRevB.99.020506}.
The strong fluctuations persist above the critical pressure, and they are
related to the superconductivity in the PM state.
This result supports the scenario that the FM fluctuations mediate the
superconductivity in UCoGe\cite{PhysRevLett.108.066403,NatComm.8.14480}.
At the intermediate pressure, the phase separation of the FM and PM phases was
observed even in the SC state.
This result is indicative of a weak first-order FM transition, as expected from
the theory of the itinerant ferromagnetism \cite{RevModPhys.88.025006}.
In the FM SC state of UCoGe at ambient pressure, the FM transition was observed
in the entire sample volume, whereas
the SC state was found to be inhomogeneous, as revealed by the two-component
nuclear spin-lattice relaxation rate $1/T_1$\cite{JPSJ.79.023707}.
Such a behavior suggests the presence of the self-induced vortex state and
the normal state at the grain boundary.
On the contrary, a recent study indicates that the SC state in UCoGe
is homogeneous when the FM phase is completely
suppressed at the pressure of 0.67\,GPa \cite{PhysRevB.99.020506}.
It is interesting to observe the emergence of superconductivity
at the intermediate pressure values, where the FM and PM states coexist
separately in the sample.
Hence, it is important to determine the precise pressure phase diagram, namely,
the FM transition line inside the SC state.
In this paper, we focus our study on the $^{59}$Co NQR measurements of UCoGe
in the critical pressure region and discuss the relationship between the
two SC states occurring in the FM state and the PM state.

\begin{figure}[t]
    \centering
    \includegraphics{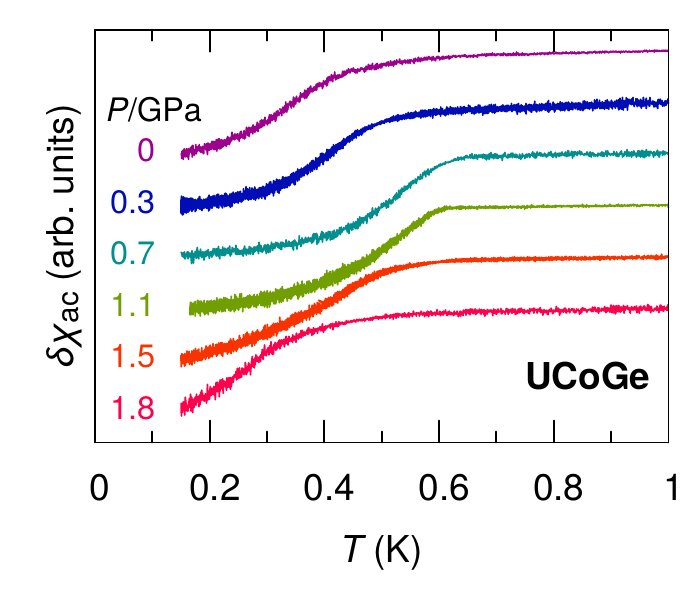}
    \caption{\label{fig:chiac}(Color online)
    Temperature dependence of ac susceptibilities of a single-crystalline
    UCoGe at several pressures measured by an NQR coil.
    The curves are shifted vertically to avoid overlapping.
    The results at 0, 0.7, and 1.1\,GPa are from a previous study
    \cite{PhysRevB.99.020506}.
    The frequency used in measurements was approximately
    in the range of 5--9\,MHz.
    }
\end{figure}

A single-crystalline UCoGe sample was used in this study.
The details of the sample properties and the measurement setup have been
previously reported elsewhere \cite{PhysRevB.99.020506,PhysRevB.100.035203}.
Hydrostatic pressure was applied using a piston-cylinder-type cell
with Daphne oil 7373 as a pressure medium.
The applied pressure was 0.3\,GPa, as determined by the SC transition
temperature of Pb.
The phase separation of the FM and PM phases persists in the SC state at this
pressure.
The FM and the SC transition temperatures of UCoGe at this pressure were
$T_\text{Curie} \sim 1.45$\,K and $T_\text{SC} = 0.50$\,K, respectively.
$T_\text{Curie}$ was determined from the $^{59}$Co NQR measurements, and
$T_\text{SC}$ was obtained from the ac susceptibility measurement
(Fig.~\ref{fig:chiac}) from an $LC$ circuit using an NQR coil.
A $^3$He-$^4$He dilution refrigerator was used for cooling the pressure cell
down to 0.15\,K\@.
The $^{59}$Co NQR ($I=7/2$) measurement was performed at zero external field.
The rf pulse field $H_1$ was applied parallel to the $c$ axis.
$1/T_1$ obtained from the $^{59}$Co NQR measurements detects the
FM fluctuations along the $c$ axis
because the nuclear quantization axis is close to the $a$ axis
\cite{PhysRevLett.105.206403} and $1/T_1$ detects the magnetic
fluctuations perpendicular to the quantized axis.
Hence, this quantity enables us to study the temperature and pressure dependence
of the FM fluctuations in details.
Since the resonant frequency depends on the internal magnetic field, $1/T_1$
can be separately measured at the FM and the PM sites when the phase separation
occurs.
The $1/T_1$ values measured in UGe$_2$ indicate
that the SC transition occurs only in the FM state\cite{JPSJ.74.2675},
which is an advantageous feature of the NQR method.

\begin{figure}
    \centering
    \includegraphics[scale=0.96]{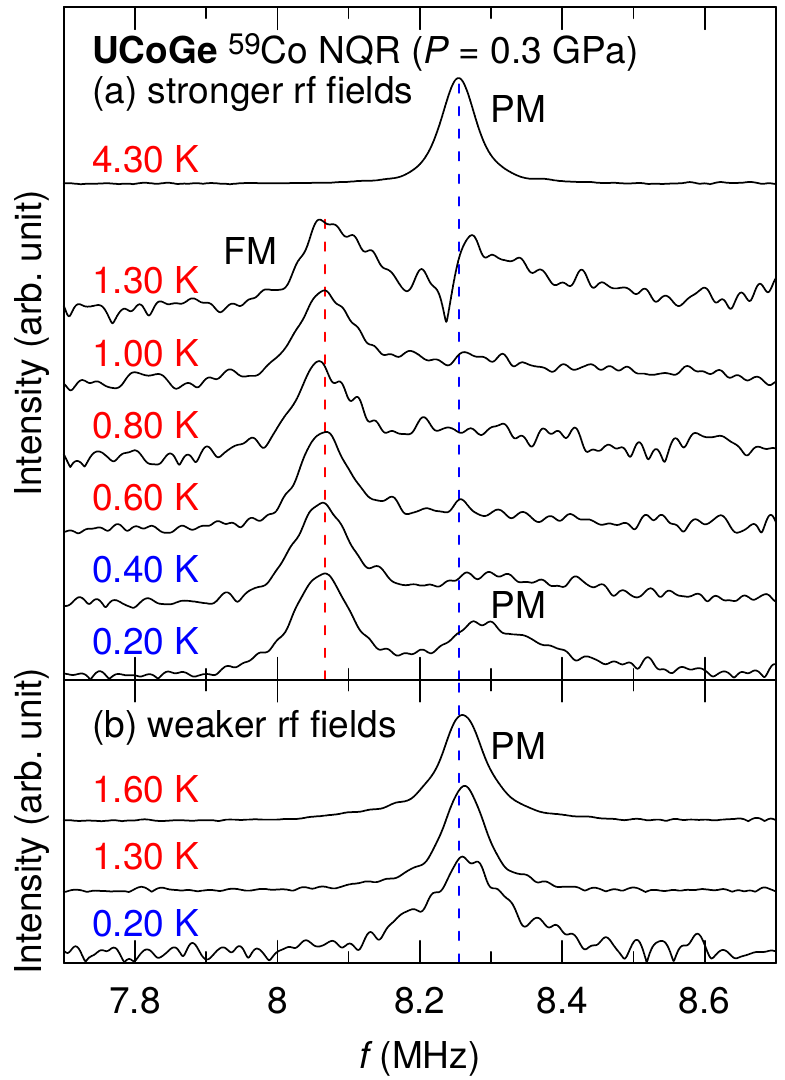}
    \caption{\label{fig:spectrum}(Color online)
    Temperature dependence of the $^{59}$Co NQR spectra at 0.3\,GPa in a
    single-crystalline UCoGe sample with (a) stronger and (b) weaker rf pulse
    fields.
    The spectra are obtained by the summation of the Fourier-transformed spin
    echoes at different frequencies.
    The rf-field intensity is 6 times stronger for the spectra in (a)
    than that in (b).
    The spectra are normalized by the peak intensity at each temperature.
    The vertical blue dashed line indicates the frequency of the PM signal at
    4.3\,K, while the red dashed line in (a) indicates the FM signal at 0.20\,K.
    The PM signal at 1.30\,K in (a) splits owing to the excess power of the
    rf pulse field for this site.
    }
\end{figure}

Figure \ref{fig:spectrum} shows the temperature dependence of the $^{59}$Co NQR
spectra around 8\,MHz arising from the $\pm 5/2 \leftrightarrow \pm 7/2$
transitions at 0.3\,GPa.
The quadrupole parameters were determined as
$\nu_\text{Q} = 2.83$\,MHz and $\eta=0.524$, which are slightly shifted compared
with the ambient-pressure values of $\nu_\text{Q} = 2.85$\,MHz and $\eta=0.52$
\cite{JPSJ.79.023707} owing to the shrinkage of the lattice under pressure.
At lower temperatures, the spectral shape depended on the intensity of the
applied NQR rf pulse field, and a stronger rf field was preferred for
observing the FM signal, as shown in Fig.~\ref{fig:spectrum}(a).
This result clearly shows the phase separation of the FM and PM phases
down to 0.20\,K, as observed in our previous NQR study
\cite{PhysRevB.99.020506}.
We confirmed that the SC anomaly is seen in the NQR spectra even with the
stronger rf pulse field, as mentioned below;
therefore, the effect of the rf heating is rather small.
Although the measurement of the FM signal above 1.30\,K was not possible
owing to the weak NQR signal intensity, $T_\text{Curie}$ was estimated to be
below 1.60\,K at 0.3\,GPa using $1/T_1$ values.
The frequency of the FM signal did not shift with a change in the temperature,
indicating that the internal magnetic field is almost
temperature-independent below $T_\text{Curie}$.
Such a behavior was also observed at the ambient pressure\cite{JPSJ.79.023707},
which is a characteristic of the first-order phase transition.
The PM signal was not clearly visible in the temperature range of 0.6--1.0\,K;
however, it was again observed in the SC state (0.20 and 0.40\,K)\@.
There are two possibilities of this recovery of the PM signal intensity
in the SC state.
One is the change in the power of the rf pulse field in the SC state as
the PM signal is more distinct with the weaker rf pulse fields
[Fig.~\ref{fig:spectrum}(b)].
The rf power becomes weaker inside the sample below $T_\text{SC}$
due to the SC diamagnetic shielding effect; hence, the PM signal becomes 
stronger in the SC state.
Another possibility may be associated with the intrinsic effects due to an
increase in the volume fraction of the PM state upon entering the
SC state, which has larger thermodynamic stability compared with the FM SC state.
We discuss the latter possibility in this article.

\begin{figure}
    \centering
    \includegraphics[scale=0.96]{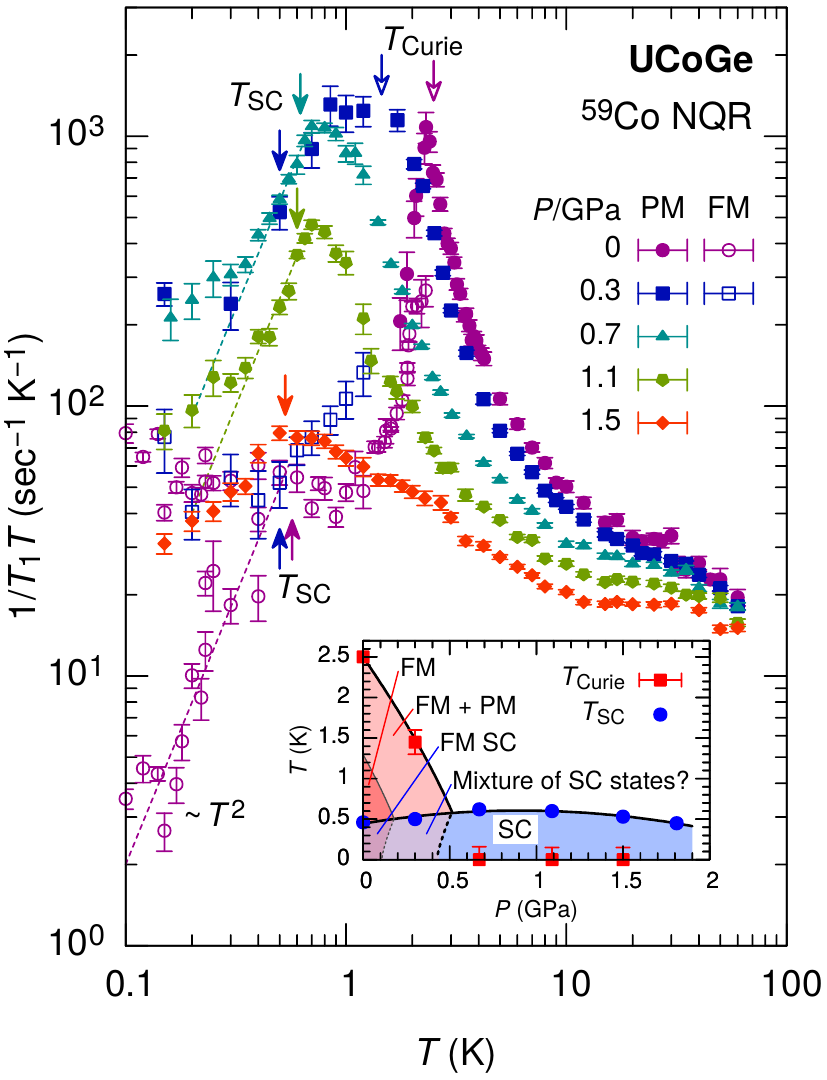}
    \caption{\label{fig:t1}(Color online)
    Nuclear spin-lattice relaxation rate divided by temperature $1/T_1T$
    obtained from the $^{59}$Co NQR measurement in a single-crystalline UCoGe.
    The data at 0\,GPa, 0.3\,GPa above 5\,K as well as 0.7, and 1.1\,GPa are
    adapted from our previous studies\cite{JPSJ.79.023707,PhysRevB.99.020506}.
    $1/T_1$ values are measured at approximately 8\,MHz.
    The solid symbols indicate the results of the PM state, whereas
    the open symbols indicate those of the FM state.
    The solid vertical arrowheads indicate $T_\text{SC}$ at each pressure
    determined by the ac susceptibility, whereas the open arrowheads indicate
    $T_\text{Curie}$ obtained by the NQR results.
    The dashed lines below $T_\text{SC}$ show the expected behavior
    in the case of a line-nodal gap ($1/T_1T \sim T^2$).
    Inset: Pressure--temperature phase diagram of the UCoGe
    sample up to 1.8\,GPa.
    The solid lines are guides to the eye.
    The dashed lines (a FM transition line inside the SC state) are not
    observed experimentally but follow a theoretical
    suggestion\cite{JPSJ.82.094711} and the experimental
    expectation from the recovery of the PM signal.
    }
\end{figure}

With the weaker rf pulse fields (typically with $1/6$ times of the intensity of
the stronger ones), the FM signal was not observed, and the PM signal was
dominant, as shown in Fig.~\ref{fig:spectrum}(b).
The signal intensity multiplied by temperature ($IT$) at 0.20\,K
decreased to $\sim$ 6\% compared with the value at 1.30\,K, and a significant
broadening of the spectrum was observed.
$IT$ is expected to be temperature-independent, if there are no phase transitions
nor magnetic fluctuations present; therefore, the observed behavior is
consistent with the presence of the strong FM fluctuations and the FM state.
The rf-field dependence of the NQR spectra indicates that the optimal rf power
differs at the PM and FM sites.
Typically, a weaker NQR rf field is preferable
when the system is close to the FM transition; hence,
the rf-field dependence may arise from the different FM fluctuations
in the FM and PM signals.
The origin of such behavior is confirmed by the $1/T_1$ measurements,
as shown in Fig.~\ref{fig:t1}.
$1/T_1$ of the FM and the PM phases were also measured with different power
of the rf pulse fields.
It is worth noting that the actual fraction of the PM and the FM phases
cannot be determined from the NQR results.

\begin{figure}
    \centering
    \includegraphics{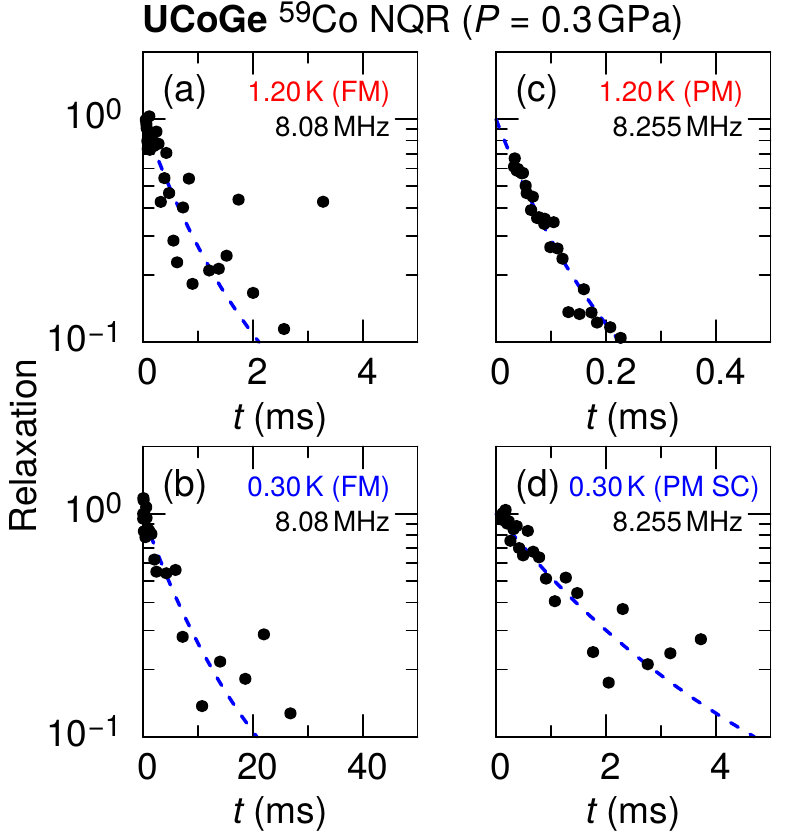}
    \caption{\label{fig:relaxation}(Color online)
    Nuclear relaxation curves of UCoGe obtained by $^{59}$Co NQR measurements
    at 0.3\,GPa in the phase-separation region.
    (a,b) Measurements at the FM site.
    (c,d) Measurements at the PM site.
    The results of the normal (1.20\,K) and the SC (0.30\,K) states are shown.
    The dashed lines are the best fit of the experimental results
    for the estimation of $1/T_1$ by a single component
    of the relaxation curve.
    Note that the horizontal scale is different for each data set.
    }
\end{figure}

Figure \ref{fig:t1} shows the temperature dependence of $1/T_1$ obtained by
the $^{59}$Co NQR measurement divided by temperature at different pressures.
The inset of Fig.~\ref{fig:t1} shows the pressure--temperature phase diagram of
the UCoGe sample.
$1/T_1$ was determined from the nuclear relaxation curve $R(t)$ obtained by
the NQR intensity at a time $t$ after applying an rf pulse field.
Some relaxation curves at 0.3\,GPa are shown in Fig.~\ref{fig:relaxation}.
The results of $1/T_1$ at 0.3\,GPa below 4.2\,K and 1.50\,GPa were obtained in
the present study, and the other data displayed in Fig.~\ref{fig:t1}
are adapted from our previous studies\cite{JPSJ.79.023707,PhysRevB.99.020506}.
At ambient pressure, $1/T_1T$ exhibited a peak at $T_\text{Curie}$ and turned
to decrease below $T_\text{Curie}$.
This peak temperature shifted to lower as the pressure increases, and it is
attributed to the suppression of the FM state.
The $1/T_1T$ peak at 0.3\,GPa is broader compared with that at
other pressure values, and the value of $1/T_1T$ of the PM
site decreased below $T_\text{SC}$ at 0.3\,GPa.
The value of $1/T_1T$ at $T_\text{SC}$ is larger than that at the ambient
pressure.
The value of $T_\text{SC}$ increased at 0.7 and 1.1\,GPa compared with
the ambient-pressure value, and a significant increase was also observed in
$1/T_1T$ at $T_\text{SC}$ in these pressures.
The FM transition was not detected at 0.7\,GPa and above.
At 1.5\,GPa, $T_\text{SC}$ decreases, and the value of $1/T_1T$ at $T_\text{SC}$
was also suppressed.
These results are in good agreement with the scenario that the FM fluctuations
are closely related to the enhancement of the superconductivity in UCoGe.

$1/T_1$ below $T_\text{Curie}$ of the FM and PM signals were measured at 0
and 0.3\,GPa.
At 0\,GPa, $1/T_1$ values of the PM and FM signals were close to each other,
whereas the $1/T_1$ at the PM site was nearly an order of magnitude faster
than that at the FM state at 0.3\,GPa.
The PM site had a strong FM fluctuations even below $T_\text{Curie}$, and
it started to decrease at approximately 0.8\,K at 0.3\,GPa, which is lower than
$T_\text{Curie} \sim 1.45$\,K and higher than $T_\text{SC}=0.50$\,K\@.
The $1/T_1$ of the FM site exhibited rather conventional behavior below
$T_\text{Curie}$ at 0.3\,GPa.

At the FM site in the SC state at 0\,GPa, two components were observed in the
nuclear relaxation curves\cite{JPSJ.79.023707}.
The longer component arises from the SC portion owing to the opening of the gap
with a line-node-like structure ($1/T_1T \sim T^2$),
whereas the shorter component exhibited a normal-metal-like behavior
($1/T_1T$ is constant).
This inhomogeneous $1/T_1$ is ascribed to the coexistence of the FM and SC
states \cite{JPSJ.79.023707}.
On the other hand, the behavior of $1/T_1$ at 0.3\,GPa in the SC state
differs from that at 0\,GPa.
A clear SC anomaly was observed in the PM signal and $1/T_1$ was determined
with a single component, which is indicative of the homogeneous SC state,
as shown in Fig.~\ref{fig:relaxation}(d).
The temperature dependence of $1/T_1T$ at 0.3\,GPa in the SC state exhibited
a residual density of states (DOS) behavior.
The behavior observed for the PM signal in this study is similar to that of the
higher-pressure PM state without the FM state\cite{PhysRevB.99.020506}.
On the contrary, a clear SC anomaly was not detected at the FM site even at
$T<T_\text{SC}$, and the nuclear relaxation curves could be fitted with a
single component.
The longer component (ascribed to the SC state) was not observed at this
pressure, as shown in Fig.~\ref{fig:relaxation}(b).
The $1/T_1T$ exhibited a normal-metal-like behavior down to 0.15\,K at the
FM site, and this behavior is similar to the fast component in the FM SC state
at 0\,GPa.
There is a possibility that a small fraction of the SC component exists in the
FM state at 0.3\,GPa; however,
the fraction of superconductivity in the FM state is smaller at
0.3\,GPa than at 0\,GPa.

The result of the $1/T_1$ measurement in the SC state at the PM signal at
0.3\,GPa suggests that the PM SC state is homogeneous.
This result implies that the phase separation between the
FM SC and PM SC states does not affect the PM SC phase.
The small SC fraction (if any) in the FM state
may be associated with the transition of a portion of the FM SC phase
into the PM SC state at $T_\text{SC}$.
The coexistence of the FM and SC states may cost an additional energy
owing to the Meissner effect,
and the fraction of the FM state may decrease below $T_\text{SC}$.
The reinforcement of the PM signal intensity in the SC state
in Fig.~\ref{fig:spectrum}(a) can be interpreted as a result of this
re-distribution of the PM and FM states, although the change in the FM
fluctuations is also important for discussion of the intensity of the NQR
spectra.
If the re-distribution occurs, the FM transition line inside the SC state
(the dashed line in the inset of Fig.~\ref{fig:t1}) is expected to have a
positive slope, which is in agreement with the theoretical
prediction\cite{JPSJ.82.094711}.
Hence, precise measurements are required to determine the volume fraction of
each phase and this phase transition line.
Furthermore, it is necessary to verify the mechanism responsible for the
reduction of the fraction of the SC state inside the FM state.

Interestingly, $1/T_1T$ of the PM SC state is faster than that
of the ``broken'' normal state at the FM site at 0.3\,GPa.
Based on the scenario that the FM fluctuations are the pairing glue for
the Cooper pairs in UCoGe,
this difference in $1/T_1T$ may lead to different $T_\text{SC}$ values
of the PM and FM phases in the phase separation region.
This scenario also leads to the re-distribution of the FM and the PM states
below $T_\text{Curie}$ of the PM state owing to the different SC
condensation energies.
To increase the condensation energy, some portion of the FM phase
transitions into the PM SC phase below $T_\text{SC}$, which results in the
lower $T_\text{SC}$ of the remaining FM state.
Although the smooth $T_\text{SC}$ values across the FM transition line have been
reported in the previous studies
\cite{JPSJ.77.073703,PhysRevLett.103.097003,PhysRevB.94.125110},
a small jump of $T_\text{SC}$ may exist owing to the weak first-order
FM transition.

Another interesting phenomenon observed in the NQR in UCoGe is the reduction
of the $1/T_1T$ value at the PM site below $T_\text{Curie}$.
Naively, the $1/T_1T$ values at the PM site are expected to increase
monotonically as the temperature decreases because the local $T_\text{Curie}$
at the PM site is always lower than the measurement temperature, and the FM
fluctuations are stronger as the temperature is close to $T_\text{Curie}$.
The reduction of $1/T_1T$ is clearly observed at 0\,GPa and is rather
ambiguous at 0.3\,GPa (Fig.~\ref{fig:t1}).
When the FM transition occurs, a phase separation also occurs between the PM and
the FM states.
Our results indicate that the PM site may be affected by the FM site,
in contrast to the behavior below $T_\text{SC}$.
It is a future task to resolve the origin of this unusual behavior.

In conclusion, we performed $^{59}$Co NQR measurements on a single-crystalline
UCoGe under 0.3\,GPa to investigate the SC state in the phase-separation
region of the FM and PM phases from the microscopic point of view.
The intensity of the NQR spectra of the PM site increased in the SC state
compared with the FM signal.
This phenomenon can be interpreted as a result of the re-distribution of the
PM and the FM states in the SC state, and is consistent with a positive slope
of the FM transition line inside the SC state in the
pressure--temperature phase diagram.
$1/T_1$ of the PM signal was homogeneous in the SC state and
exhibited a clear drop owing to the opening of the SC gap at 0.3\,GPa.
On the contrary, $1/T_1$ of the FM site did not exhibit a clear SC anomaly below
$T_\text{SC}$.
This suggests that the SC state in the FM phase is weaker than that in
the PM phase owing to the Meissner effect in the FM state.

\begin{acknowledgment}
The authors would like to thank H. Kotegawa, H. Tou,
Y. Tokunaga, T. Hattori, D. Aoki, Y. Maeno,
S. Yonezawa, A. Daido, Y. Yanase, J.-P. Brison, D. Braithwaite, A. Pourret,
C. Berthier, A. de Visser, J. Flouquet, and V. P. Mineev for their valuable
discussions.
One of the authors (MM) is a Research Fellow of
Japan Society for the Promotion of Science (JSPS).
This work was supported by the Kyoto University LTM Center,
Grants-in-Aid for Scientific Research
(No.~JP15H05745 and No.~JP17K14339),
Grants-in-Aid for Scientific Research on Innovative Areas ``J-Physics''
(No.~JP15H05882, No.~JP15H05884, and No.~JP15K21732), and Grants-in-Aid
for JSPS Research Fellows (No.~JP17J05509 and No.~JP19J00336) from JSPS.
\end{acknowledgment}

\bibliographystyle{jpsj}
\bibliography{bibliography}

\end{document}